\documentclass[apj,iop]{emulateapj}

\usepackage{graphicx}
\usepackage{amsmath, amssymb}

\begin{document}

\title{Determination of Stochastic Acceleration 
	Model Characteristics in Solar Flares}
\author{Qingrong Chen and Vah\'e Petrosian}
\affil{Department of Physics, Stanford University, Stanford, CA 94305, USA}

\shorttitle{Determination of Stochastic Acceleration Model Characteristics in Solar Flares}
\shortauthors{Chen \& Petrosian}

\begin{abstract}

Following our recent paper \citep{Petrosian10},
we have developed an inversion method to determine
the basic characteristics of the particle acceleration mechanism
directly and non-parametrically from observations 
under the leaky box framework. 
In the above paper, we demonstrated this method for obtaining the energy
dependence of the escape time.
Here, by converting the Fokker-Planck equation to its integral form,
we derive the energy dependences of the energy diffusion coefficient and
direct acceleration rate for stochastic acceleration
in terms of the accelerated and escaping particle spectra.
Combining the regularized inversion method of 
\citet{Piana07} and our procedure, 
we relate the acceleration characteristics in solar flares directly to 
the count visibility data from {\it RHESSI}.
We determine the timescales for electron escape, pitch angle scattering,
energy diffusion, and direct acceleration at the loop top acceleration region
for two intense solar flares
based on the regularized electron flux spectral images.
The X3.9 class event shows dramatically different energy dependences
for the acceleration and scattering timescales,
while the M2.1 class event shows a milder difference.
The M2.1 class event could be consistent with 
the stochastic acceleration model with a very steep turbulence spectrum.
A likely explanation of the X3.9 class event could be that
the escape of electrons from the acceleration region is not governed by
a random walk process,
but instead is affected by magnetic mirroring,
in which the scattering time is proportional to the escape time
and has an energy dependence similar to the energy diffusion time.

\end{abstract}

\keywords{acceleration of particles --- Sun: flares --- Sun: X-rays, gamma rays}

\maketitle

\section{Introduction}
Solar flares are a complex multiscale phenomenon 
powered by the explosive energy release from non-potential magnetic fields 
through reconnection in the solar corona.
The total energy released in a large flare can reach 
up to $\sim$$10^{32}$--$10^{33}$ erg within $\sim$$10^{2}$--$10^{3}$ s
and $\sim$10--50\% of this energy goes into acceleration of electrons and ions 
to relativistic energies in the impulsive phase 
\citep{Lin76, Lin03, Emslie12}.
In particular, the suprathermal electrons produce 
hard X-ray (HXR) emission up to a few hundred keV through the well understood 
bremsstrahlung process \citep{Lin74, Dennis88, Krucker08a, Holman11}. 

HXR observations in the past two decades 
from the {\it Yohkoh}/Hard X-ray Telescope and
the Reuven Ramaty High Energy Solar Spectroscopic Imager 
\citep[{\it RHESSI};][]{Lin02, Hurford02} have significantly 
advanced our understanding of electron acceleration in solar flares.
Detection of distinct coronal HXR sources 
located near the top of the flare loop 
in addition to the commonly seen footpoint (FP) sources
\citep[e.g.,][]{Masuda94, Aschwanden02, Petrosian02, BB06, 
Krucker08b, Ishikawa11b, ChenQ12, Simoes13}
has revealed that the primary electron acceleration takes place in the corona 
with an intimate relation to the energy release process 
by magnetic reconnection.
More recently {\it RHESSI} further observed
a second coronal X-ray source located above the loop top (LT) source.
The higher energy emission of the two coronal sources is found 
to be closer to each other 
\citep[e.g.,][]{Sui03, Sui04, LiuW08, ChenQ12, LiuW13}.
This property, 
complemented with the extreme ultraviolet (EUV) observations 
of the context \citep[e.g.,][]{WangT07},
further suggests that electron acceleration occurs most likely
in the reconnection outflow regions, rather than in the current sheet
\citep[][]{Holman12, LiuW13}.

Several different acceleration mechanisms may operate in the outflow regions
as a result of the explosive energy release by reconnection, 
involving either the kinetic effects of 
small amplitude electromagnetic fluctuations 
or the spatial-temporal variations of the large scale magnetic fields.
Among these mechanisms, the model of stochastic acceleration (SA),
also known as the second-order Fermi process \citep{Fermi49},
has achieved considerable success in interpreting the high energy features
of solar flares \citep[e.g.,][]{Miller97, Petrosian12}.
Resonant interactions of particles 
with a broad spectrum of plasma waves or turbulence in the corona,
presumably excited by the large scale outflows from the reconnection region,
lead to momentum diffusion and pitch angle scattering of particles
\citep{Sturrock66b, Tsytovich66, Tsytovich77, Tverskoi67, Tverskoi68}.
Several variants of the SA mechanism
have been applied to acceleration of electrons and ions in solar flares 
\citep[e.g.,][]{Melrose74, Barbosa79, 
Ramaty79, Ryan91, Hamilton92, Steinacker92, Miller95, Miller96, Park97,
Petrosian04, Emslie04a, LiuS06, Grigis06, Bykov09, Bian12, Fleishman13}.
Resonant pitch angle scattering,
a necessary prerequisite for efficient acceleration 
\citep[e.g.,][]{Tverskoi67, Miller97b, Melrose09},
increases the time electrons stay at the LT acceleration region
\citep[e.g.,][]{Petrosian99}, 
before they escape to the thick target FPs of the flare loop.
This enhances the HXR radiation at the coronal LT region
and naturally explains the aforementioned HXR morphological 
structure associated with the flare loop.

Particle spectra resulting from SA by turbulence are generally described 
by the so-called leaky box version of the Fokker-Planck kinetic equation 
\citep[e.g.,][]{Ramaty79, Steinacker92, Park95, Petrosian04}. 
The accelerated and escaping electron spectra and 
the resulting bremsstrahlung HXR spectra at the LT and FPs 
are found to be sensitive to the turbulence spectrum
and the background plasma properties \citep{Petrosian99, Petrosian04}. 
There have been continued efforts to constrain 
the wave-particle interaction coefficients and the property of turbulence 
from solar flare HXR (and $\gamma$-ray) observations,
mainly through a parametric forward fitting procedure
\citep{Hamilton92, Park97, LiuW09b}.
Although there has been systematic theoretical modeling of the SA mechanism
in attempt to explain the spectral features of {\it RHESSI} HXR observations
\citep{Petrosian04, Grigis06},
the spatially resolved imaging spectroscopic data 
from {\it RHESSI} have been rarely under direct quantitative comparison 
to constrain the SA model characteristics.

By taking advantage of the recently developed electron flux spectral images
\citep{Piana07} via regularized inversion from the 
{\it RHESSI} count visibility data,
\citet{Petrosian10} initiated direct determination of
the SA model characteristics from the radiating electron flux spectra
at the LT and FP regions.
We have derived 
the energy dependences of the escape time and pitch angle scattering time.
In this paper, 
by fully utilizing the leaky box Fokker-Planck equation 
describing the acceleration process, 
we further derive the energy dependence of the energy diffusion coefficient,
which also gives the direct acceleration rate by turbulence,
directly and non-parametrically from 
the spatially resolved electron spectra in solar flares.
This provides a complete determination of all unknown SA model quantities 
in the Fokker-Planck equation.

In the next section, we present the equations 
describing the particle acceleration, transport, and radiation processes.
In Section \ref{sec:Determination}, 
we show the general determination of the escape time and 
how the inversion of the Fokker-Planck kinetic equation 
leads to the energy diffusion coefficient 
in terms of purely observable quantities. 
In Section \ref{sec:RHESSI},
we apply the formulas to two {\it RHESSI} solar flares
to determine the SA model characteristics based on the electron flux images.
In the final section we give a brief summary and discuss implications of 
the results for acceleration and transport of electrons in solar flares.
 
\section{Acceleration, Transport, and Radiation}

In this paper, we are interested in the spatially averaged
characteristics of the mechanism accelerating the background thermal particles. 
For a homogeneous acceleration region these would give the actual values. 
In application to solar flares, the acceleration region
with volume $V$, cross section ${\cal A}$, and size $L=V/{\cal A}$, 
which we assume to be consisted of a fully ionized hydrogen plasma
with background number density $n_{\rm LT}$, 
would be embedded at the apex of the flare loop. 
We define a free streaming time across the acceleration region as 
$\tau_{\rm cross}=L/v$.
We assume that this region contains a certain level of turbulence
to scatter and accelerate particles.

\subsection{Leaky Box Acceleration Model}
\label{sec:LeakyBox}

The very complex details of particle diffusion in the momentum space
due to wave-particle interactions
are most commonly illuminated by the quasilinear theory
\citep[][and references therein]{Kennel66, Schlickeiser89},
through the momentum and pitch angle diffusion coefficients,
namely $D_{pp}$, $D_{p\mu}$, and $D_{\mu\mu}$.
However, acceleration by turbulence (and some other mechanisms, e.g., shocks) 
requires a pitch angle scattering time ($\tau_{\rm scat}\sim 1/D_{\mu\mu}$) 
that is much shorter than other timescales.
As a result, particles rapidly attain a nearly isotropic distribution,
and instead of free streaming, 
they diffuse out of the accelerator via a random walk process. 
By translating 
the spatial diffusion into an escape term from the accelerator,
and transforming from the momentum space to the energy domain,
the evolution of the particle distribution function $N(E,t)$,
averaged over the pitch angle and integrated over the physical space,
is conventionally described by the leaky box model.

We use the following slightly modified variant of the leaky box Fokker-Planck 
equation% 
\footnote{\label{fnote:FPEq}
From the standard form of the Fokker-Planck formalism \citep{Chandrasekhar43}, 
$\frac{\partial}{\partial t}N=
\frac{\partial^2}{\partial E^2} \left[D_{\rm EE}N\right]
-\frac{\partial}{\partial E}\left[A_d(E)N\right]$,
where $D_{\rm EE}\equiv \langle\frac{(\Delta E)^2}{2\Delta t}\rangle$
and $A_d(E)\equiv \langle\frac{\Delta E}{\Delta t}\rangle
=A(E) + \frac{d D_{\rm EE}}{dE}$ for SA only,
it is easy to show that the total energy of the accelerated particles 
${\cal E}(t)=\int_0^\infty EN(E, t)dE$ varies with time as
$\frac{d}{dt}{\cal E} = \int_0^\infty A_d(E)N dE$. 
Thus, it is $A_d(E)$, rather than $A(E)$, 
that gives the actual energy gain rate or direct acceleration rate 
\citep{Tsytovich66, Tsytovich77, Ramaty79}.
Insertion of $A_d(E)$ into the above equation yields 
a form of Equation (\ref{eq:FPEq}),
the steady state of which is a first-order (instead of second-order)
ordinary differential equation for $D_{\rm EE}$.}
\citep[][]{Park96, Petrosian12},
which is more convenient for our purpose here,
\begin{equation}
\frac{\partial N}{\partial t}=
\frac{\partial}{\partial E}\left[D_{\rm EE}\frac{\partial N}{\partial E}\right]
-\frac{\partial}{\partial E}\left[(A-\dot{E}_{\rm L})N\right] 
-\frac{N}{T_{\rm esc}}+\dot{Q},
\label{eq:FPEq}
\end{equation}
where $D_{\rm EE}$ and $A(E)$ are 
the energy diffusion coefficient and the acceleration rate 
(due to turbulence and all other interactions), respectively, 
$\dot{E}_{\rm L}(E)$ is the energy loss rate, and 
$\dot{Q}(E)$ and $N(E)/T_{\rm esc}(E)$ are the rates of injection of 
seed particles and escape of the accelerated particles, respectively. 

The energy diffusion coefficient by turbulence is
\begin{equation}
D_{\rm EE}= v^2 D(p)\equiv \frac{v^2}{2}\int_{-1}^{1} 
\left(D_{pp}-\frac{D_{p\mu}^2}{D_{\mu\mu}} \right)d\mu.
\end{equation}
If turbulence is the only agent of acceleration, 
then the acceleration rate is \citep{Petrosian12}
\begin{equation}
A(E)=\frac{D_{\rm EE}}{E}\xi(E),\ 
{\rm with}\ \xi(E)=\frac{2\gamma^{2}-1}{\gamma^2+\gamma},
\label{eq:AEDE}
\end{equation}
where $\gamma$ is the Lorentz factor.
The escape time is related to the spatial diffusion of particles 
along the magnetic field lines, 
which depends on the pitch angle scattering time $\tau_{\rm scat}$ as 
\citep[e.g.,][]{Schlickeiser89, Steinacker92, Petrosian12} 
\begin{equation}
T_{\rm esc} = \frac{\tau_{\rm cross}^2}{\tau_{\rm scat}},
\ {\rm with}\ 
\tau_{\rm scat} =\frac{1}{8}
\int_{-1}^1 \frac{(1-\mu^2)^2}{D_{\mu\mu}} d\mu.
\label{eq:Kss}
\end{equation}
The above relation is valid when the scattering time is
much shorter than the crossing time.
We further add $\tau_{\rm cross}$ to the escape time,
\begin{equation}
T_{\rm esc}\simeq \tau_{\rm cross}+ \frac{\tau_{\rm cross}^2}{\tau_{\rm scat}},
\label{eq:Tesc}
\end{equation}
which extends its validity to the opposite case and 
assures that the escape time is longer than the crossing time.
For further discussions about the above equations, 
see \citet{Petrosian04} and \citet{Petrosian12}.
The effect of the geometry of the large scale magnetic fields
on the escape time will be discussed in Section \ref{sec:Summary}. 

For solar flare X-ray radiating electrons below a few MeV, 
the energy loss rate $\dot{E}_{\rm L}$ is dominated by 
Coulomb collisions with the background electrons,
\begin{equation}
\dot{E}_{\rm L} = \dot{E}_{\rm L}^{\rm Coul}=
4\pi r_0^2 m_{\rm e} c^4 n\ln\Lambda/v,
\label{eq:ECoul}
\end{equation}
where $n$ is the background electron density,
$r_0$ is the classical electron radius with $4\pi r_0^2= 10^{-24}$ cm$^2$, and
$\ln\Lambda$ is the Coulomb logarithm taken to be 20 for solar flare conditions.

In most astrophysical systems, in particular in solar flares,
the dynamic timescale is generally much longer than 
the acceleration and other timescales,
then it is justified to treat the steady state leaky box equation.
Solution of this equation provides the spectrum and the escape rate
of the accelerated particles, $N(E)$\ and $N(E)/T_{\rm esc}$,
respectively, or equivalently, the accelerated and escaping flux spectra
(in units of particles cm$^{-2}$ s$^{-1}$ keV$^{-1}$),
\begin{align}
F_{\rm acc}=\frac{vN}{V},
F_{\rm esc}=\frac{N}{{\cal A}T_{\rm esc}}=
\left(\frac{\tau_{\rm cross}}{T_{\rm esc}}\right)F_{\rm acc}.
\label{eq:Fesc}
\end{align}
From the above particle spectra, we can obtain the escape time as
\begin{equation}
T_{\rm esc}=\left(\frac{F_{\rm acc}}{F_{\rm esc}}\right)\tau_{\rm cross},
\label{eq:Tesc0}
\end{equation}
and from Equations (\ref{eq:Tesc} and \ref{eq:Kss}), 
we can obtain the pitch angle scattering time $\tau_{\rm scat}$ and
the averaged pitch angle diffusion rate $\langle D_{\mu\mu} \rangle$. 

\subsection{Particle Transport}
\label{sec:Transport}

The escape rate $N(E)/T_{\rm esc}$ serves as the seed source $\dot{Q}^{\rm tr}$
for the subsequent transport of particles outside the acceleration region. 
If the particles lose all their energy in the transport region, 
i.e., we are dealing with a thick target process, 
then the volume integrated particle spectrum is governed by
the steady state transport kinetic equation \citep{Longair92},
\begin{equation}
\frac{\partial N^{\rm tr}}{\partial t} = 
\frac{\partial }{\partial E}\left(\dot{E}_{\rm L}^{\rm tr}N^{\rm tr}\right) + 
\dot{Q}^{\rm tr}=0,
\label{eq:Thicktarget}
\end{equation}
where $\dot{E}_{\rm L}^{\rm tr}$ is the energy loss rate
at the thick target transport region.
Then solution of this equation gives rise to the effective 
thick target radiating particle spectrum \citep{Longair92, Johns92},
\begin{equation} 
N_{\rm eff}^{\rm tr}(E) = \frac{1}{\dot{E}_{\rm L}^{\rm tr}} 
\int_E^{\infty} \frac{N}{T_{\rm esc}} dE,
\label{eq:Neff}
\end{equation}
For Coulomb collisions in solar flares, 
the energy loss rate $\dot{E}_{\rm L}^{\rm tr}$ 
should be evaluated with the mean density $n_{\rm tr}$
from the loop legs to FPs.

\subsection{Bremsstrahlung HXR Radiation}
\label{sec:Brem}

In solar flares, the accelerated and escaping electrons produce 
bremsstrahlung HXR emission along the flare loop,
for which the angle-averaged differential photon flux
(in units of photons s$^{-1}$ keV$^{-1}$) 
is written as a linear Volterra integral equation of the first kind,
\begin{equation}
J(\epsilon)
=\int_{\epsilon}^\infty X(E) \sigma(\epsilon, E) dE,
\label{eq:Je}
\end{equation}
where $\sigma(\epsilon, E)$ is the angle-averaged bremsstrahlung cross section.
The quantity $X(E)$ represents the integration over the volume of interest
of the electron flux spectrum $F(E,s)$
multiplied with the background proton density $n(s)$,
\begin{equation}
X(E)\equiv \int n(s)F(E,s) {\cal A}(s) ds,
\label{eq:XE}
\end{equation}
where ${\cal A}(s)$ is the cross section of the loop 
along the magnetic field lines.
In what follows, we refer to $X(E)$ as the volume integrated 
radiating electron flux spectrum.
Thus at the LT acceleration region,
\begin{equation}
X_{\rm LT}(E)=n_{\rm LT}V F_{\rm acc}= n_{\rm LT} vN(E).
\label{eq:XLT}
\end{equation}

The transport of the escaping electrons from the loop legs to FPs
is described by the classical thick target model
\citep{Brown71, Syrovat-Skii72, Petrosian73}.
The radiating electron flux spectrum 
integrated over the whole thick target, but mainly at the FPs,
produced by the escaping electrons is given by \citep[e.g.,][]{Park97}
\begin{equation}
X_{\rm FP}(E)= 
n_{\rm tr}vN_{\rm eff}^{\rm tr}(E),
\label{eq:XFP}
\end{equation}
where $N_{\rm eff}^{\rm tr}$ is given by Equation (\ref{eq:Neff}). 
It should be noted that as a result of the density dependence
of the Coulomb energy loss rate, 
the thick target radiating electron spectrum $X_{\rm FP}$ 
and consequently the bremsstrahlung photon spectrum $J_{\rm FP}$
are independent of the thick target density profile
\citep[e.g.,][]{Syrovat-Skii72, Park97}.

As explained below, the volume integrated radiating electron flux spectra 
$X_{\rm LT}$ and $X_{\rm FP}$ 
can be obtained directly and non-parametrically from {\it RHESSI} data.

\section{Determination of Model Quantities}
\label{sec:Determination}

The two unknown diffusion coefficients 
in the SA model are $D(p)$ and $D_{\mu\mu}$, 
or equivalently, $D_{\rm EE}$ and $T_{\rm esc}$,
which we aim to determine from observations.
In comparison,
one has relatively good knowledge or estimate of 
the energy loss rate $\dot{E}_{\rm L}$ and the source term $\dot{Q}$,
which depend primarily on the background medium properties.
Therefore, given the accelerated and escaping particle flux spectra
$F_{\rm acc}$ and $F_{\rm esc}$ from observations,
in particular, $X_{\rm LT}$ and $X_{\rm FP}$ from solar flares,
we can in principle determine the two unknown model quantities.

\subsection{Escape Time}

As already indicated above, 
one can in general determine the first unknown quantity,
namely, the escape time, simply from the ratio between 
$F_{\rm acc}$ and $F_{\rm esc}$ (Equation \ref{eq:Tesc0}),
or alternatively from the ratio between $N(E)$ and $N(E)/T_{\rm esc}$.
For solar flare bremsstrahlung, on the other hand, 
we deal with a thick target transport process 
and the escaping electrons produce an effective
radiating spectrum $N_{\rm eff}^{\rm tr}$.
%radiating outside the acceleration region 
%$N_{\rm eff}^{\rm tr}$ is given by Equation (\ref{eq:Neff}). 
By differentiating Equation (\ref{eq:Neff}), 
we then determine the escape time as \citep[see also][]{Petrosian10} 
\begin{equation}
T_{\rm esc}=\frac{E}{\dot{E}_{\rm L}^{\rm tr}}
\frac{N}{N_{\rm eff}^{\rm tr}}
\left(-\frac{d\ln N_{\rm eff}^{\rm tr}}{d\ln E}
-\frac{d\ln \dot{E}_{\rm L}^{\rm tr}}{d\ln E}\right)^{-1}.
\label{eq:Tesc1}
\end{equation}
Note that for Coulomb collisions in a cold target, we have 
$\frac{d\ln \dot{E}_{\rm L}^{\rm tr}}{d\ln E} 
= -\frac{1}{\gamma^2+\gamma} \simeq -\frac{1}{2}$
at the non-relativistic limit.

\subsection{Energy Diffusion Coefficient}
Now we are left with the second unknown quantity, 
namely, the energy diffusion coefficient, 
and it turns out that the derivation for $D_{\rm EE}$ is very simple.
By using the relation between $A(E)$ and $D_{\rm EE}$ 
(Equation \ref{eq:AEDE}), 
we rewrite the steady state leaky box equation as below,
\begin{equation}
\frac{d}{d E}
\left[D_{\rm EE}\left(\frac{d N}{d E}-\frac{N}{E}\xi \right)\right]
+\frac{d}{d E}\left(\dot{E}_{\rm L}N\right)
=\frac{N}{T_{\rm esc}}-\dot{Q},
\end{equation}
Integration of the above equation from $E$ to $\infty$ gives 
\begin{equation}
D_{\rm EE}=E
\left[ \dot{E}_{\rm L}+
\frac{1}{N}
\int_E^\infty \left(\frac{N}{T_{\rm esc}} - \dot{Q}\right)dE\right]
\left(\xi- \frac{d \ln{N}}{d \ln{E}} \right)^{-1}.
\label{eq:DEE}
\end{equation}
For particle energies far above the injection energy,
acceleration results in $N/T_{\rm esc} \gg \dot{Q}$,
so that $\dot{Q}$ can be ignored from the above equation. 
Therefore, 
given the escape time $T_{\rm esc}$ as determined above,
we can derive the formula for $D_{\rm EE}$ once again
purely in terms of observables.

This formula can be further simplified for the thick target transport model. 
The integral inside the square brackets is related to 
the effective thick target radiating spectrum for the escaping particles 
(Equation \ref{eq:Neff}). 
Thus we express $D_{\rm EE}$ as 
\begin{equation}
D_{\rm EE}= E\dot{E}_{\rm L} \left( 1 + 
\frac{\dot{E}_{\rm L}^{\rm tr}N_{\rm eff}^{\rm tr}}{\dot{E}_{\rm L}N} \right)
\left(\xi- \frac{d \ln{N}}{d \ln{E}} \right)^{-1}.
\label{eq:DEE1}
\end{equation} 

In summary, by differentiating the effective thick target radiating spectrum 
due to the escaping particles and converting the Fokker-Planck equation
to the integral form, we can express the unknown model quantities,
the escape time $T_{\rm esc}$ and 
the energy diffusion coefficient $D_{\rm EE}$, 
purely in terms of observables with minimal assumptions.

\subsection{Solar Flare Radiating Electron Spectra}

For Coulomb collisional energy loss in solar flares, we have 
$\dot{E}_{\rm L}^{\rm tr}/\dot{E}_{\rm L}=n_{\rm tr}/n_{\rm LT}$ 
and the following relation,
\begin{equation}
\frac{\dot{E}_{\rm L}}{\dot{E}_{\rm L}^{\rm tr}}\frac{N}{N_{\rm eff}^{\rm tr}}
= \frac{n_{\rm LT}N}{n_{\rm tr}N_{\rm eff}^{\rm tr}}
= \frac{X_{\rm LT}}{X_{\rm FP}}.
\label{eq:Ratio}
\end{equation}
Thus, in terms of the volume integrated radiating electron flux spectra
$X_{\rm LT}$ and $X_{\rm FP}$ in solar flares,
we rewrite Equation (\ref{eq:Tesc1}) for the escape time as
\begin{equation}
T_{\rm esc}=
\tau_{\rm L}\left(\frac{X_{\rm LT}}{X_{\rm FP}}\right)
\left(-\frac{d\ln X_{\rm FP}}{d\ln E}+\frac{2}{\gamma^2+\gamma}\right)^{-1},
\label{eq:Tesc2} 
\end{equation}
and Equation (\ref{eq:DEE1}) for the energy diffusion coefficient as% 
\footnote{The relations
$\frac{d\ln v^2}{d\ln E} = \frac{2}{\gamma^2+\gamma}$ and
$\xi+\frac{d\ln v}{d\ln E} = \frac{2\gamma}{\gamma+1}$ are used.}
\begin{equation}
D_{\rm EE}= 
\frac{E^2}{{\tau}_{\rm L}}
\left(1 + \frac{X_{\rm FP}}{X_{\rm LT}}\right)
\left(-\frac{d \ln{X_{\rm LT}}}{d \ln{E}} + 
\frac{2\gamma}{\gamma+1}\right)^{-1},
\label{eq:DEE2} 
\end{equation}
where $\tau_{\rm L} = {E}/{\dot{E}_{\rm L}}$ is the energy loss time 
at the LT acceleration region (with density $n_{\rm LT}$). 
We further define the energy diffusion time 
due to turbulence as $\tau_{\rm diff}={E^2}/{2 D_{\rm EE}}$ 
and direct acceleration time as $\tau_{\rm acc}={E}/{A_d}$ 
(see Footnote \ref{fnote:FPEq}).

\subsection{Interplay between Timescales}
The shape of the accelerated electron spectrum is 
a result of the interplay between the competing processes 
involved in the leaky box Fokker-Planck Equation (\ref{eq:FPEq}).
Conversely, we can gain some insight into these physical processes 
from the electron spectra as we have shown above.
Both $T_{\rm esc}$ and $D_{\rm EE}$ 
primarily depend on the ratio $X_{\rm LT}/X_{\rm FP}$.
By eliminating $X_{\rm LT}/X_{\rm FP}$ from 
Equations (\ref{eq:Tesc2} and \ref{eq:DEE2}),
we relate the timescales for the physical processes as below,
\begin{equation}
\frac{1}{\tau_{\rm diff}} = 
\frac{2}{\eta_{\rm LT}}
\left(\frac{1}{\tau_{\rm L}} + \frac{1}{\eta_{\rm FP}T_{\rm esc}} \right),
\label{eq:timescales}
\end{equation}
where $\eta_{\rm LT} = 
-\frac{d \ln{X_{\rm LT}}}{d \ln{E}} + \frac{2\gamma}{\gamma+1}$
and $\eta_{\rm FP}
=-\frac{d\ln X_{\rm FP}}{d\ln E}+\frac{2}{\gamma^2+\gamma}$.
If the (non-relativistic) X-ray radiating electron spectra
$X_{\rm LT}$ and $X_{\rm FP}$ in solar flares are nearly power laws,
then both $\eta_{\rm LT}$ and $\eta_{\rm FP}$ vary very slowly with energy.

We now consider two extreme cases.
On the one hand, if $X_{\rm LT}/X_{\rm FP}\ll 1$,
which is applicable to most flare observations, then we have 
$\eta_{\rm FP}T_{\rm esc}\ll \tau_{\rm L}$ and roughly 
$\tau_{\rm acc} \sim \tau_{\rm diff} \simeq 
(\eta_{\rm LT}\eta_{\rm FP}/2)T_{\rm esc}$.
On the other hand, if $X_{\rm LT}/X_{\rm FP}\gg 1$,
which is representative for 
a few very rare events with an extremely bright LT source, then we have 
$\eta_{\rm FP}T_{\rm esc}\gg \tau_{\rm L}$ and
$\tau_{\rm acc} \sim \tau_{\rm diff}\simeq (\eta_{\rm LT}/2)\tau_{\rm L}$.

\section{Applications to {\it RHESSI} Observations}
\label{sec:RHESSI}

The volume integrated radiating electron flux spectra 
$X_{\rm LT}$ and $X_{\rm FP}$ have been generally inferred from
the HXR spectra of the LT and FP sources using 
the Volterra integral Equation (\ref{eq:Je}),
which is an ill-posed inverse problem and there are no unique solutions. 
This is commonly carried out by a forward fitting procedure 
\citep[e.g.,][]{Holman03}, 
but there have also been attempts to determine these electron flux spectra
by the inversion of this equation \citep{Brown06}. 
Several methods, such as analytic solution \citep{Brown71}, 
matrix inversion \citep{Johns92}, 
and regularized inversion \citep{Piana03, Kontar05}
have been used for this task. 
Here we use the more recent and direct procedure described below.

\subsection{Regularized Electron Imaging Spectroscopy}

\citet{Piana07} noted that
the most fundamental product of the temporal modulation from {\it RHESSI}
as a Fourier imager is the count {visibilities},
the Fourier components of the source spatial distribution, 
which are related via essentially the same Volterra Equation (\ref{eq:Je}) 
to the electron flux visibilities,
the Fourier components of the so-called electron flux spectral images.
By the same regularized inversion method as mentioned above,
\citet{Piana07} first inverted the electron flux visibility spectrum 
from the count visibility spectrum. 
This requires knowledge of the bremsstrahlung cross section and the 
detector response function.
Then by applying visibility-based imaging algorithms to the these visibilities, 
they reconstructed the images of the mean radiating electron flux 
multiplied by the column depth ${\cal N}(x, y)$ along the line-of-sight,% 
\footnote{More exactly, the electron flux spectral images represent 
$a^2{\cal N}(x,y)F(x, y; E)/10^{50}$, 
where $x$ and $y$ are in units of arcsec and 
$a=7.25\times 10^{7}$ cm arcsec$^{-1}$.
From these images, 
the sum of the pixel intensities within one region of interest,
after multiplication by the square of the pixel size (in units of arcsec),
yields the volume integrated radiating electron flux spectra 
$X(E)$ defined in Equation (\ref{eq:XE}) for that region,
in units of 10$^{50}$ electrons cm$^{-2}$ s$^{-1}$ keV$^{-1}$.
In the current paper, we have corrected our misinterpretation 
of the observed electron flux spectra by up to a constant 
as made in the upper panel of Figure 2 in \citet{Petrosian10}.}
namely, $X(x,y;E)={\cal N}(x,y)F(x, y; E)$, 
where $x$ and $y$ are the spatial coordinates.
From these electron flux images over a sequence of energy bins, 
one can then extract the volume integrated radiating electron flux spectra
$X(E)=\int X(x,y,E)dxdy$
for spatially separated LT and FP sources of solar flares.
With availability of this regularized ``electron" imaging spectroscopy, 
one can now better constrain the acceleration and transport processes 
in solar flares \citep[e.g.,][]{Prato09, Petrosian10, Torre12, GuoJ13,
Massone13, Codispoti13}.
\citet{Torre12}, 
assuming a spectrum of accelerated electrons, 
used a similar integration of the transport equation
to determine the energy loss rate along the flare loop.

As can be explicitly seen from Equations (\ref{eq:Tesc2} and \ref{eq:DEE2}), 
simultaneous detection of both the LT and FP sources in solar flares
over a wide energy range is essential to determine 
the SA model characteristics as a function of electron energy. 
For this purpose, we have carried out 
a systematic search of high energy events \citep{ChenQ09}, 
for which both the LT and FP emission
during the impulsive phase is imaged by {\it RHESSI}.
We have found a few such events close to the solar limb
with the HXR emission detected 
above 50 keV from both the LT and FP sources.

We reconstruct the regularized electron flux images
using the MEM\_NJIT algorithm \citep{Schmahl07}.
In the data analysis performed below, 
the electron flux spectra $X_{\rm LT}$ and $X_{\rm FP}$
are extracted from the electron images using the Object Spectral Executive
\citep[OSPEX; ][]{Smith02} package of the Solar SoftWare.
The fittings in this paper are implemented using 
a non-linear least squares fitting program, MPFIT, 
based on the Levenberg-Marquardt algorithm \citep{Markwardt09, More77}.

We apply the above data analysis procedure 
to the {\it GOES} X3.9 class solar flare on 2003 November 3 and 
the {\it GOES} M2.1 class flare on 2005 September 8
and determine the SA model characteristics.
In Table \ref{table_numbers}, 
we list the basic information of these two flares, and
the power law indices for the radiating electron flux spectra
and the SA model quantities and related timescales. 

\begin{table}[t]
\center
\caption{Basic information and 
power law indices of the electron flux spectra ($\propto E^{-\delta}$)
and the SA model quantities and timescales ($\propto E^{s}$) 
in two {\it RHESSI} flares.}
\begin{tabular}{lll}
Date & 2003 November 3 & 2005 September 8 \\
\hline
{\it GOES} Class & X3.9 Class & M2.1 Class \\
SOL Locator & SOL2003-11-03T09:43 & SOL2005-09-08T16:49 \\ 
Location & AR 10488 (N08, W77) & AR 10808 (S10, E81) \\
\hline
{\it RHESSI} ID & 3110316 & 5090832 \\
Duration & 09:49:34--09:50:02 UT & 16:59:40--17:00:40 UT\\
Energy & $\leq$250 keV & $\leq$130 keV \\
\hline
$X_{\rm LT}$ & $2.99 \pm 0.03$ & $4.77 \pm 0.06$ \\
$X_{\rm FP}$ & $2.45 \pm 0.05$$^{a}$
& $3.48 \pm 0.03$ \\
\hline
$T_{\rm esc}$ & $0.83 \pm 0.10$ & $0.21 \pm 0.15$ \\
$\tau_{\rm scat}$ & $-1.82 \pm 0.13$ & $-0.90 \pm 0.24$ \\
$D_{\mu\mu}$ & $1.82 \pm 0.13$ & $0.90 \pm 0.24$ \\
\hline
$\tau_{\rm diff}$ & $1.06 \pm 0.12$ & $0.52 \pm 0.15$ \\
$\tau_{\rm acc}$ & $0.96 \pm 0.15$$^{b}$ & 
		 $0.45 \pm 0.21$$^{b}$ \\
$D_{\rm EE}$ & $0.99 \pm 0.11$ & $1.49 \pm 0.17$\\ 
\hline
\end{tabular}
\tablecomments{$^a$The mean value of the broken power law fitting.}
\tablecomments{$^b$To calculate $\tau_{\rm acc}$,
we approximate the logarithmic derivative of $D_{\rm EE}$
with its power law fitting.} 
\label{table_numbers}
\end{table}

\begin{figure*}
\centering
\includegraphics[scale=0.9]{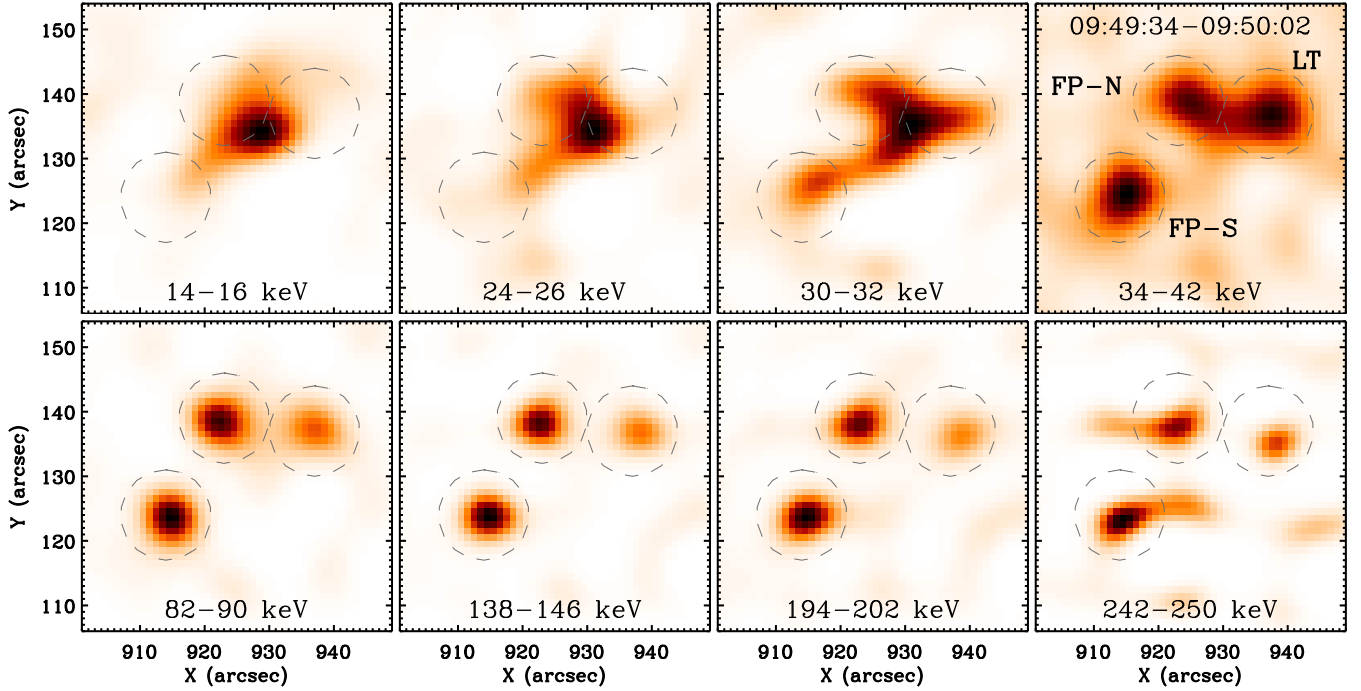}
\caption{Electron flux spectral images up to 250 keV
in the X3.9 class solar flare on 2003 November 3
reconstructed by the MEM\_NJIT method from 
the regularized electron flux visibilities. 
The images indicate one distinct LT source and 
two FP sources during the impulsive phase.
The LT source is located near the cusp structure as shown at low energies.
The three circles (dash) denote the LT and FPs.}
\label{fig:Nov03_images}
\end{figure*}

\begin{figure*}
\centering
\includegraphics[scale=0.85]{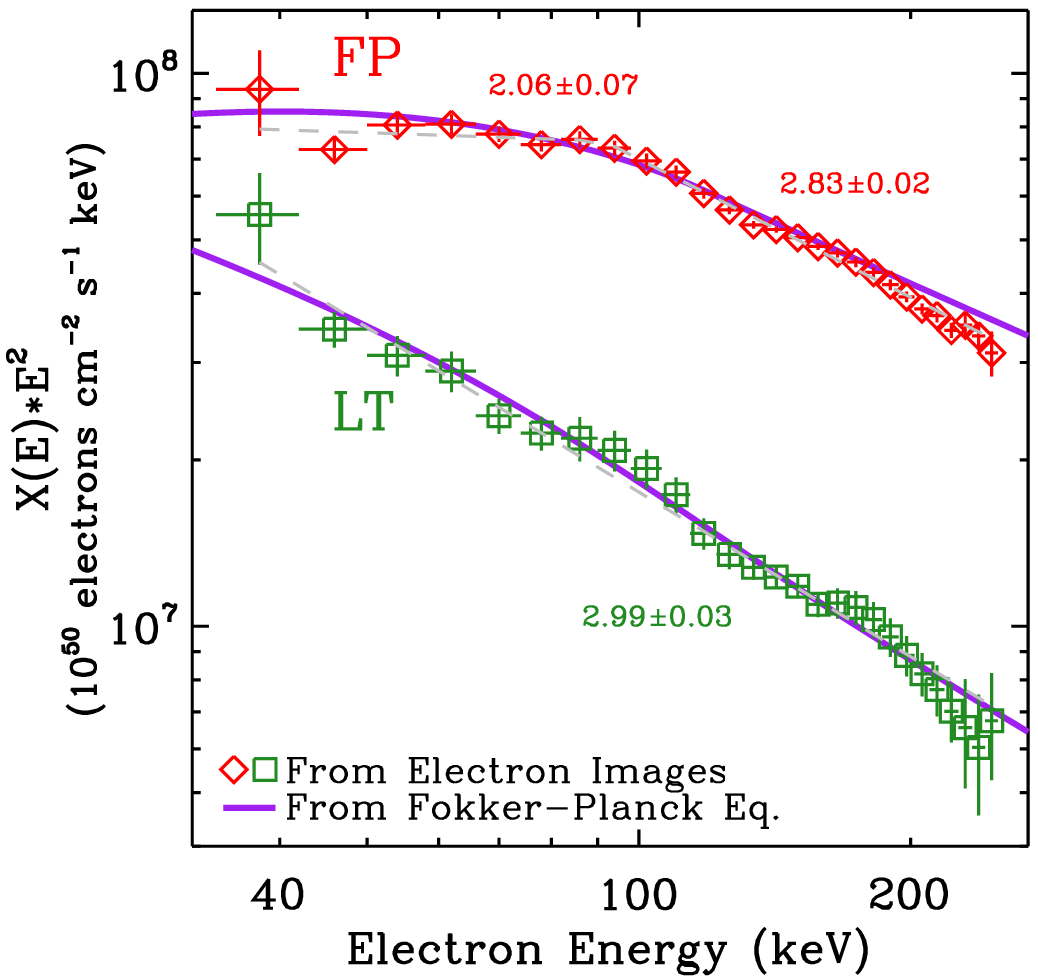}
\includegraphics[scale=0.85]{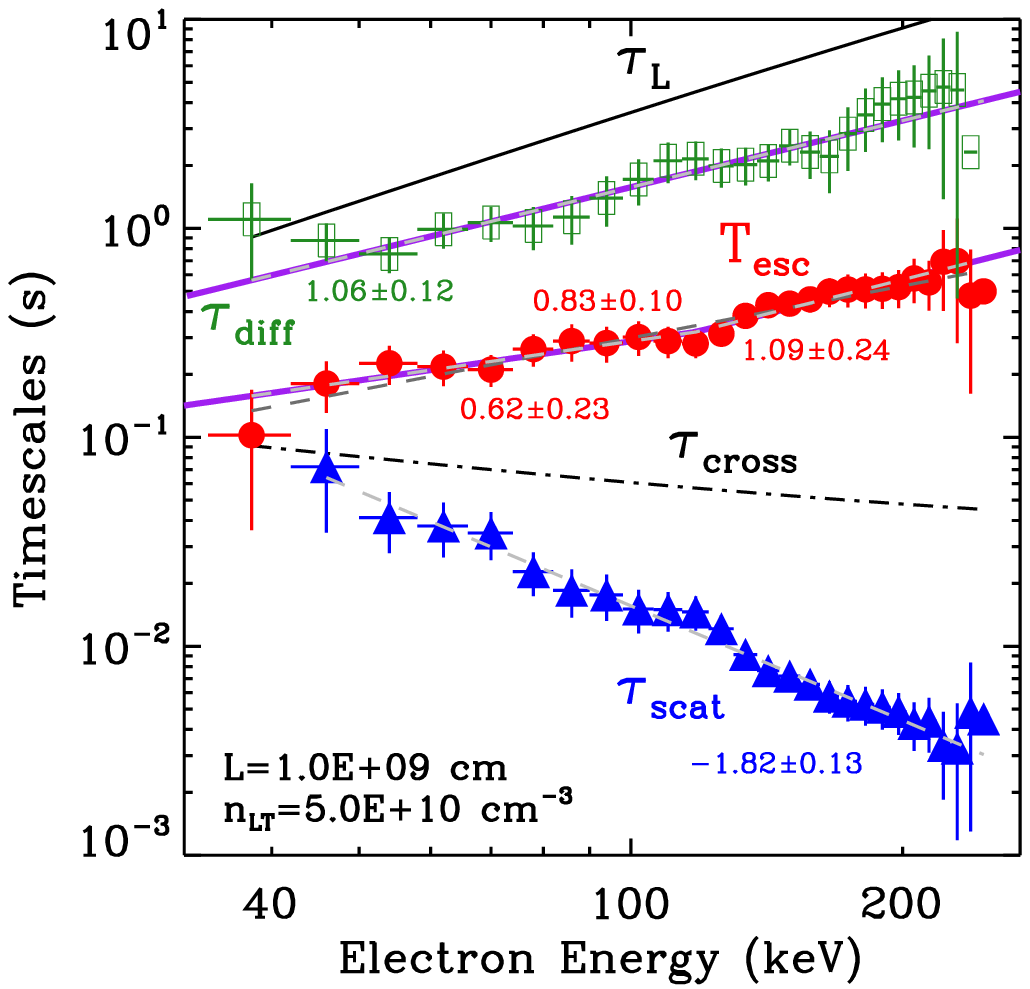}
\caption{Radiating electron flux spectra and SA model timescales
in the X3.9 class solar flare on 2003 November 3.
Left: Radiating electron flux spectra $X(E)$
from the LT region (square, green) and the FP regions summed (diamond, red),
which can be fitted by a single and a broken power law (dash, gray),
respectively.
Right: Timescales for electron escaping ($T_{\rm esc}$, circle, red),
pitch angle scattering due to turbulence 
($\tau_{\rm scat}$, triangle, blue),
energy diffusion ($\tau_{\rm diff}$, standing bar, green), 
crossing ($\tau_{\rm cross}$, dash-dot, black), 
and Coulomb energy loss ($\tau_{\rm L}$, solid, black)
at the LT acceleration region with
density $n_{\rm LT}=5\times 10^{10}$ cm$^{-3}$ and size $L=10^{9}$ cm.
The gray dash lines show 
the single or broken power law fitting of the timescales and
the numbers near these lines are the power law indices.
Furthermore, as a consistency check, 
the SA model timescales $T_{\rm esc}$ and $\tau_{\rm diff}$
(right panel, solid, purple) are used as input to
the steady state leaky box Equation (\ref{eq:FPEq}).
The accelerated electron spectrum solved numerically from this equation
and the thick target radiating electron spectrum due to the escaping 
electrons (left panel, solid, purple) exhibit very good match with 
the observed $X_{\rm LT}$ and $X_{\rm FP}$ spectra.}
\label{fig:Nov03_spectra}
\end{figure*}

\subsection{The 2003 November 3 Event}

The 2003 November 3 solar flare of X3.9 class
(Solar Object Locator: SOL2003-11-03T09:43)
is an intense solar eruptive event close to the west solar limb.
The unusually bright HXR emission from the coronal LT source, 
detectable up to 
100--150 keV along with two FP sources by {\it RHESSI} \citep{ChenQ12},
makes this event particularly suitable for our purpose 
to determine the SA model characteristics.

Figure \ref{fig:Nov03_images} shows 
the regularized electron flux spectral images. 
The LT and FP sources are clearly visible up to 250 keV, about twice 
the highest photon energy for the LT source.
We then extract the volume integrated radiating electron flux spectra 
$X(E)$ above 34 keV from the LT source and the two FP sources
(Figure \ref{fig:Nov03_spectra}, left panel).
The LT spectrum can be fitted by a power law with an index $\sim$3.0, 
while the flatter FP spectrum can be better fitted by a broken power law
with the indices $\sim$2.1 and $\sim$2.8 
below and above the break energy $\sim$91 keV, respectively.

From analysis of the X-ray images \citep{ChenQ12},
we obtain the density at the LT acceleration region 
to be $n_{\rm LT}\sim 5\times 10^{10}$ cm$^{-3}$
with the LT size to be $L\sim 10^9$ cm.
In Figure \ref{fig:Nov03_spectra} (right panel),
we plot the electron escape time and the energy diffusion time 
as calculated from the above spatially resolved spectra 
$X_{\rm LT}$ and $X_{\rm FP}$ (left panel). 
Except for the leftmost data point at the lowest energy,
the escape time $T_{\rm esc}$ is clearly much longer than 
the crossing time $\tau_{\rm cross}$
and is $\sim$5--15 times shorter than the energy loss time $\tau_{\rm L}$.
The escape time shows an overall trend increasing with energy,
and can be fitted with a power law, $T_{\rm esc} \propto E^{0.8}$.
From the escape time, we calculate the pitch angle scattering time, 
which decreases with energy as $\tau_{\rm scat} \propto E^{-1.8}$.
This implies a pitch angle diffusion rate of $D_{\mu\mu} \propto E^{1.8}$.
Here we attribute the above scattering time purely to turbulence.
Contribution from Coulomb collisions will be 
at the scale of the energy loss time 
and therefore negligible for electrons above 34 keV in this event.
The energy diffusion time varies as $\tau_{\rm diff} \propto E^{1.1}$ and 
is about half the energy loss time.
Thus we have the energy diffusion coefficient $D_{\rm EE}\propto E^{0.9}$.
The direct acceleration time $\tau_{\rm acc}$ 
is very close to the energy diffusion time.
It is obvious that the energy diffusion time and pitch angle scattering time
have very different energy dependences in this event.

\begin{figure*}[ht]
\centering
\includegraphics[scale=0.9]{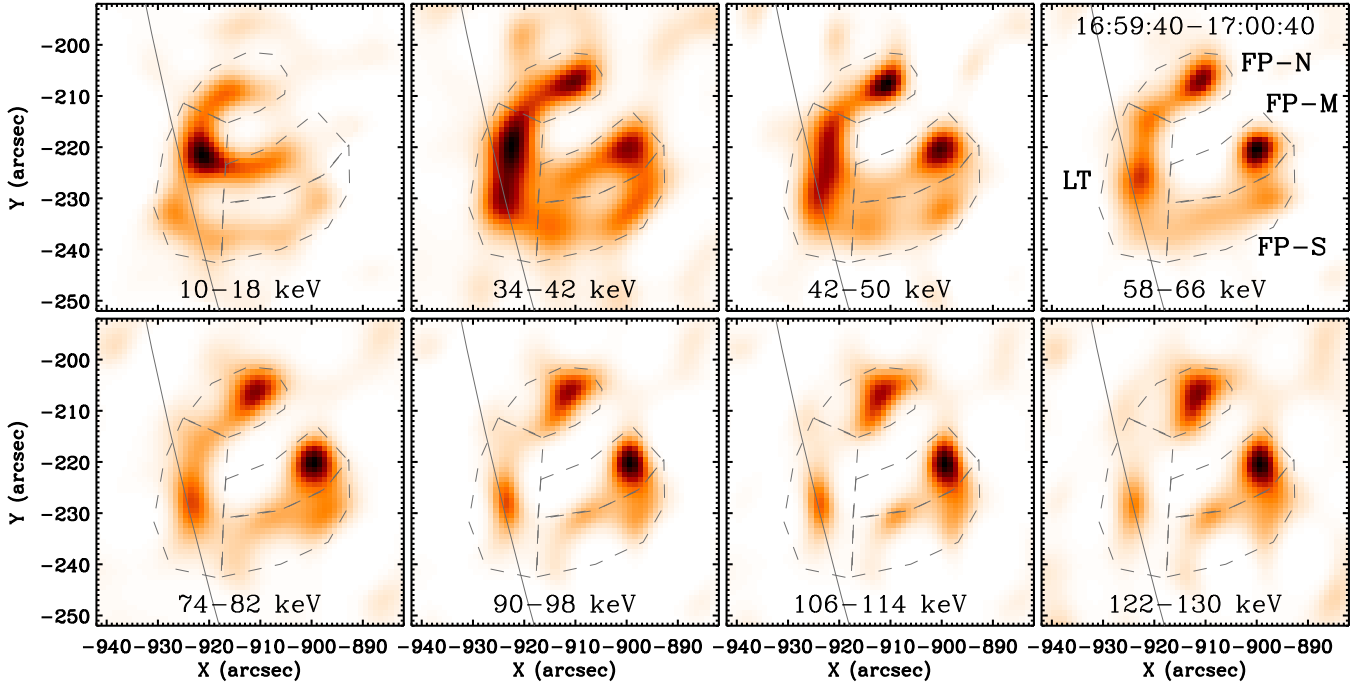}
\caption{Electron flux spectral images up to 130 keV
in the M2.1 class solar flare on 2005 September 8.
The images indicate two interacting loops.
The polygons (dash) denote the LT and FP regions
of the two loops. The solid lines shows the solar limb.}
\label{fig:Sep08_images}
\end{figure*}

\begin{figure*}[ht]
\centering
\includegraphics[scale=0.85]{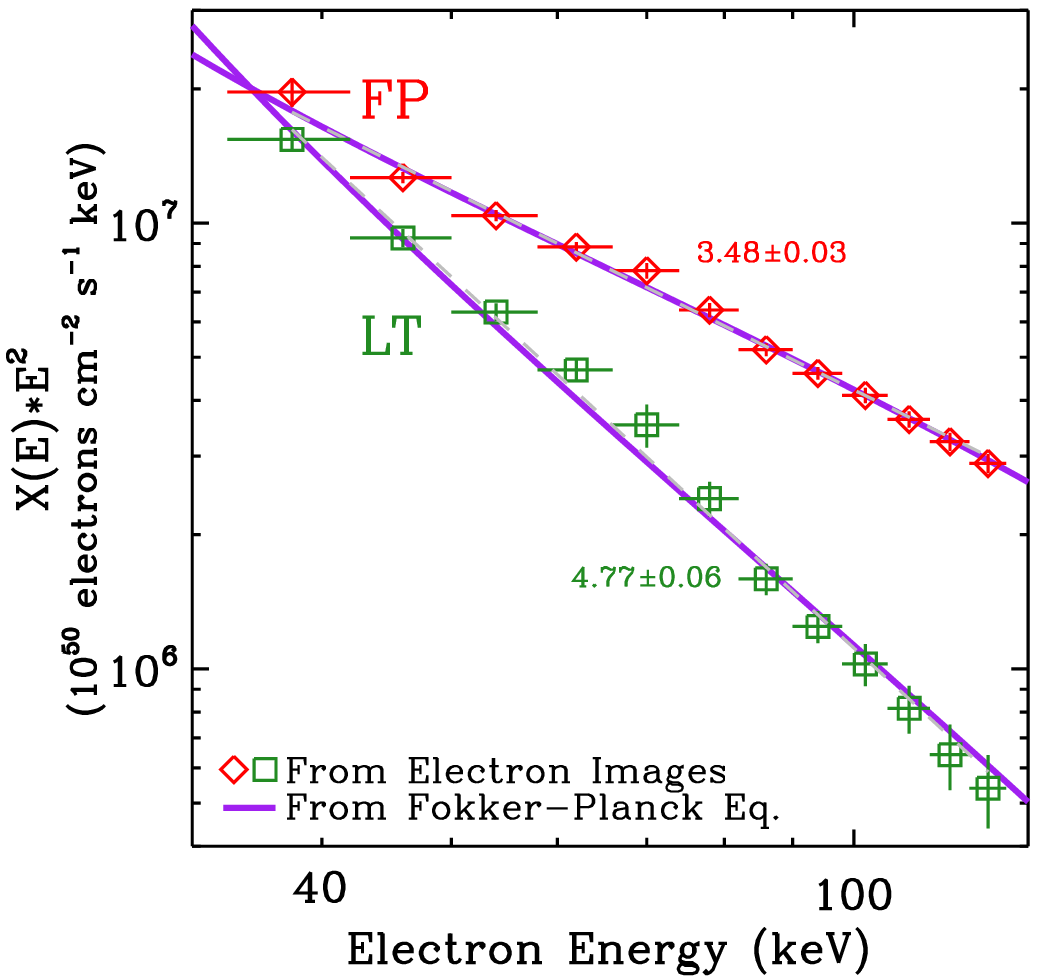}
\includegraphics[scale=0.85]{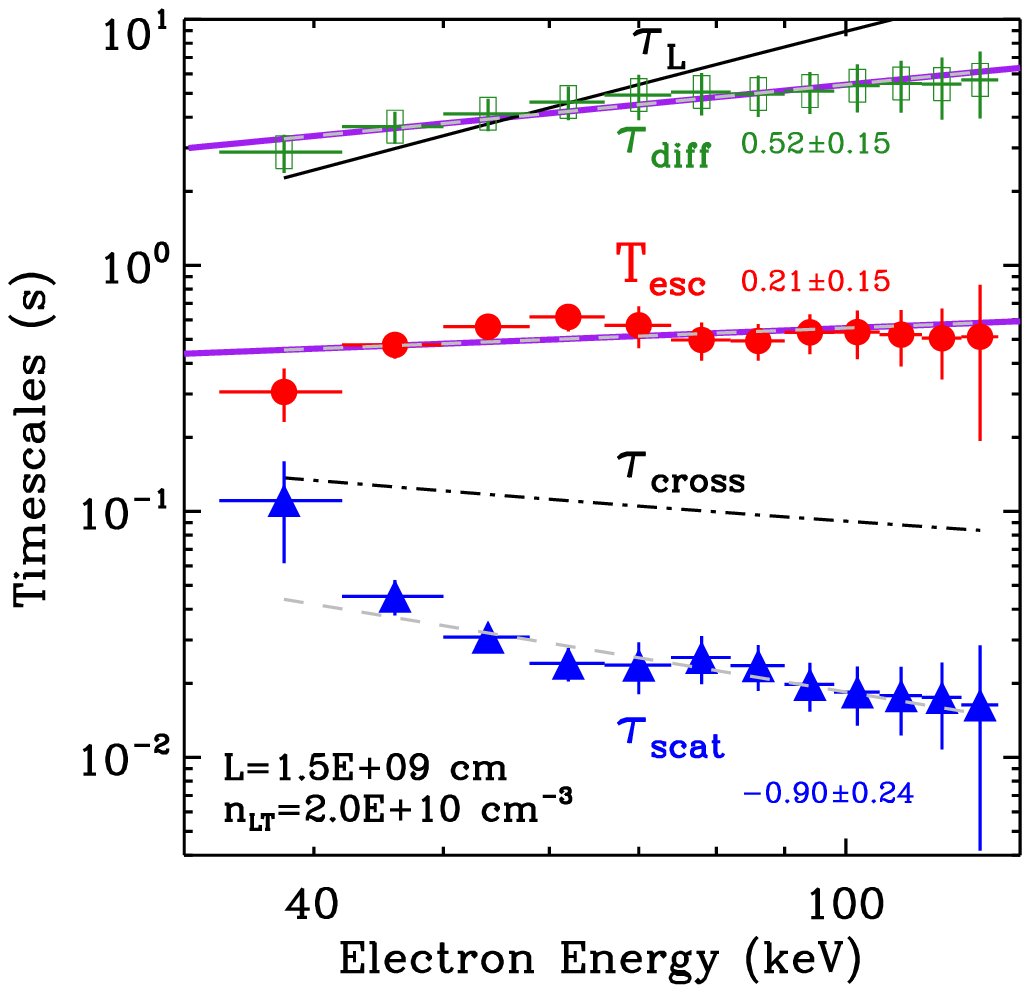}
\caption{Same as Figure 2,
but for the M2.1 class solar flare on 2005 September 8.
}
\label{fig:Sep08_spectra}
\end{figure*}

\subsection{The 2005 September 8 Event}

The 2005 September 8 solar flare (SOL2005-09-08T16:49) is an M2.1 class 
event occurring at the southeast quadrant of the Sun near the limb.
As seen from the {\it RHESSI} HXR images and 
the Transition Region and Coronal Explorer ({\it TRACE}) 
171 \AA\ EUV images, the flare consists of two interacting loops, 
with their northern loop legs visually overlapped along the line-of-sight.
Furthermore, the coronal LT source appears higher at altitude 
with increasing HXR energy \citep{ChenQ09}.
Here we model the acceleration region associated with the two loops 
as a single leaky box for the whole flare.
We take the density and size of this single accelerator to be 
$2\times 10^{10}$ cm$^{-3}$ and $1.5\times 10^9$ cm, respectively.

Figure \ref{fig:Sep08_images} 
displays the electron flux images up to 130 keV,
in which two flare loops can be clearly resolved. 
We extract the radiating electron flux spectra 
at the LT and FP sources summed over the two loops. 
As shown in Figure \ref{fig:Sep08_spectra} (left panel),
both the LT and FP spectra can be well fitted by a power law,
with the indices $\sim$4.8 and $\sim$3.5, respectively,
the difference of which is larger than that in the 2003 November 3 flare.
Due to the relatively softer LT source in this event,
all the model timescales are flatter than 
those in the 2003 November 3 flare.
As in Figure \ref{fig:Sep08_spectra} (right panel), 
the escape time can be fitted with a power law, 
$T_{\rm esc} \propto E^{0.2}$.
The scattering time varies as $\tau_{\rm scat}\propto E^{-0.9}$,
and the pitch angle diffusion rate as $D_{\mu\mu}\propto E^{0.9}$.
The energy diffusion time and acceleration time can be fitted 
with a similar power law, 
$\tau_{\rm diff} \propto \tau_{\rm acc} \propto E^{0.5}$.
The energy diffusion coefficient is found to be 
$D_{\rm EE} \propto E^{1.5}$. 
Again, the energy dependences for the energy diffusion time and
the pitch angle scattering time are very different,
but now the difference is smaller than that in the 2003 November 3 flare.

\subsection{Numerical Verification}

For the above two events, we also use the power law forms of 
the escape time $T_{\rm esc}$ 
and the energy diffusion coefficient $D_{\rm EE}$ 
determined directly from observations as input to 
the steady state leaky box Fokker-Planck Equation ({\ref{eq:FPEq}}),
and solve for the electron spectra $N(E)$ numerically 
using the Chang-Cooper finite difference scheme \citep{Chang70, Park96}.
We then calculate the effective thick target radiating spectra
for the escaping particles.
As shown in Figures \ref{fig:Nov03_spectra} and \ref{fig:Sep08_spectra}, 
these numerical model spectra in general 
agree very well with the observed spectra from {\it RHESSI}.
This is a mere self-consistency check justifying our procedure.

\section{Summary and Discussions}
\label{sec:Summary}

Following our earlier paper \citep{Petrosian10}, 
we have developed a new method for the determination of 
the energy dependences of basic characteristics of the SA mechanism. 
The particle spectrum is determined by three such characteristics, 
but only two of them, 
namely the momentum and pitch angle diffusion coefficients 
($D_{pp}$ and $D_{\mu\mu}$), play a major role. 
As is well known, for a nearly isotropic particle distribution at
a homogeneous acceleration region, 
the diffusion equation in the momentum space reduces to the so-called
leaky box Fokker-Planck equation, including 
the energy diffusion coefficient and direct acceleration rate
($D_{EE}$ and $A_d(E)$ related to $D_{pp}$) 
and escape time ($T_{\rm esc}$ related to $D_{\mu\mu}$). 
We show that 
from the observed spectra of the accelerated and escaping particles, 
we can determine these two unknowns of the SA mechanism by 
inversion of the particle acceleration and transport equations
and thus gain insight into the properties of the required plasma waves or
turbulence. 
It should be noted that, in contrast to the usual forward fitting
method, this is a non-parametric method of relating the acceleration
coefficients directly to observables. 

In this paper we have applied the above procedure to acceleration
of electrons in solar flares based on {\it RHESSI} HXR observations, 
assuming mainly SA by turbulence with negligible contribution
to acceleration from electric fields or shocks.%
\footnote{In a companion paper (V. Petrosian \& Q. Chen, in preparation),
we explore the application of this method to
acceleration of electrons in supernova remnant shocks.}
Here we also employ the regularized inversion procedure of 
\cite{Piana07} to produce the electron flux spectral images 
from the {\it RHESSI} count visibilities, 
{\it thus relating the acceleration model coefficients directly and 
non-parametrically to the raw {\it RHESSI} data.} 

We have applied our method to two intense flares observed by {\it RHESSI}
on 2003 November 3 (X3.9 class) and 2005 September 8 (M2.1 class).
The results from both events exhibit some interesting behaviors, 
as summarized below. 

$\bullet$ We find that electrons 
stay at the acceleration region much longer than 
the free crossing time and shorter than the energy loss time
\citep[see also][]{Petrosian10, Simoes13}. 
Furthermore, {\it the escape time increases with energy.} 
This is our most robust result
and is independent of the details of the acceleration mechanisms. 
The only assumption is that electrons are accelerated at the LT region and 
lose most of their energy by Coulomb collisions at the FPs. 
The twofold effects of a long escape time, which 
increases the acceleration efficiency and suppresses the escaping rate,
can naturally explain the relatively flat accelerated electron spectrum 
at the LT source, especially for the 2003 November 3 flare.

$\bullet$ A short scattering time is a possible explanation of 
this observation
and would justify the assumption of the pitch angle isotropy. 
Our results indicate that 
Coulomb scattering is not efficient enough to produce this effect, 
thus scattering by
turbulence is the most likely mechanism as advocated in the SA model.

$\bullet$ If we assume that the pitch angle scattering is 
the cause of the long escape time (Assumption I), 
then using the simple random walk relation, we find 
{\it a scattering time that decreases relatively rapidly with energy} 
and is much shorter than the crossing time
(and all other times), as required in this scenario.

$\bullet$ If we also assume the SA model (Assumption II),
in which $D_{\rm EE}$ and $A(E)$ are related by Equation (\ref{eq:AEDE}),
then we find that {\it the energy diffusion time increases with energy}
and is roughly parallel to the escape time 
for both flares (except in the few low energy bins). 
The same is true for the direct acceleration time.
This behavior is what is expected when the FP sources are stronger than 
the LT source ($X_{\rm FP} > X_{\rm LT}$).

Clearly knowledge thus gained from observations
can then be used to directly compare with theoretical model predictions. 
It should, however, be emphasized that
the derivation of the electron escape time 
does not involve assumptions about any specific acceleration mechanisms 
and thus it may impose the most severe constraint on the theoretical models.

\subsection{Comparison with SA Modeling}

We now discuss whether the above results, specifically 
the discordant energy dependences of the acceleration and scattering time, 
can be reconciled with the predictions of the SA model. 
This is a test of the above Assumption II while keeping Assumption I.

\begin{figure}
\centering
\includegraphics[scale=0.83]{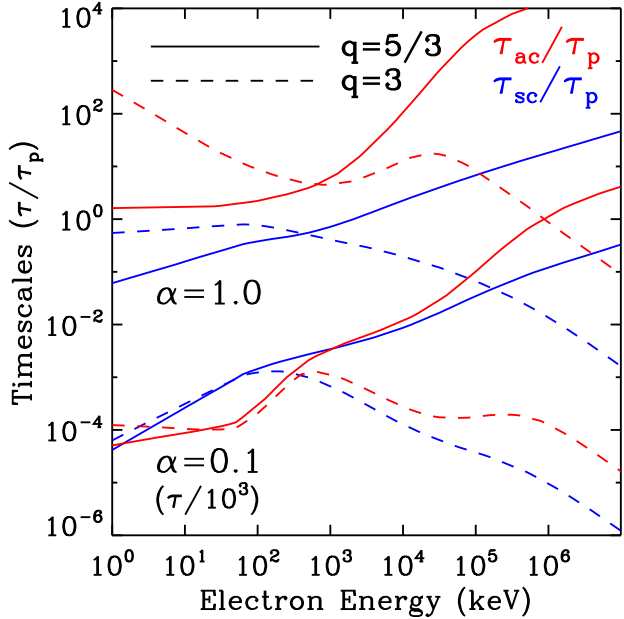}
\caption{Energy dependences of acceleration time and scattering time for 
two values of the turbulence spectral index $q$ (5/3, solid; and 3, dash)
and two values of $\alpha$ (1.0 and 0.1),
as extracted from \citet[][Figures 12 and 13 therein]{Pryadko97}.
The timescales are normalized with $\tau_p$, 
a typical timescale in turbulent plasmas.
The constant $\alpha=\omega_{pe}/\Omega_{e}$ is defined to be
the ratio of electron plasma frequency to gyrofrequency.
The timescales for $\alpha=0.1$ are shifted downward three decades 
for display.
Note that both $\tau_{\rm ac}$ and $\tau_{\rm sc}$ 
in this figure \citep[][Equations 30 and 31 therein]{Pryadko97}
are eight times shorter than 
the energy diffusion time $\tau_{\rm diff}$ (non-relativistic) and 
the pitch angle scattering time $\tau_{\rm scat}$
defined in the current paper, respectively.}
\label{fig:PP97Times}
\end{figure}

As described in the quasilinear theory,
the momentum and pitch angle diffusion coefficients
are related to the turbulence energy density spectrum ${\cal W}(k) \propto
k^{-q}$, where $k$ is the wave number,
plasma dispersion relation $\omega(k)$, and resonance conditions
\citep[e.g.,][]{Schlickeiser89, Dung94}.
In the relativistic range, we expect similar energy dependences for 
the momentum diffusion and pitch angle scattering timescales, 
$\tau_{\rm diff}\sim (c/v_A)^2\tau_{\rm scat}\propto E^{2-q}$,
where $v_A$ is the Alfv\'en speed.
In the non-relativistic regime of interest here, 
the relation becomes more complex. 
Most past works in the literature modeling the SA mechanism in solar flares
assume plasma waves propagating parallel to the large scale magnetic fields 
\citep[e.g.,][]{Steinacker92, Dung94, Pryadko97, Pryadko98, Petrosian04}. 
As shown in Figure \ref{fig:PP97Times}, there is a considerable variation 
of the model timescales at low energies, 
especially for the energy diffusion (or direct acceleration) time. 
The scattering time from the models 
increases or is nearly constant with energy
at the non-relativistic regime. 
It generally obeys an approximate relation,
$\tau_{\rm scat}\propto E^{(3-q)/6}$ \citep{Petrosian04}, 
so that a steep turbulence spectrum with $q>3$ 
is needed for a scattering time that decreases with energy. 
Thus, for the two flares studied above, 
we need $q>14$ and $q>8$, respectively.
However, for such steep turbulence spectra, 
the diffusion and acceleration timescales 
will most likely decrease with energy. 
\citet {Petrosian04} provided another approximate expression,
$\tau_{\rm diff}\propto E^{(7-q)/6}$, 
which is valid in a limited non-relativistic range. 
This would disagree with 
our results under the random walk approximation.

We therefore conclude that the observational results from 
the intense X3.9 class flare on 2003 November 3
is not consistent with the SA model predictions as presented above.
While the results from the weaker M2.1 class flare on 2005 September 8
with weaker and softer LT emission 
have less severe disagreement with the SA model if electrons 
are interacting with a steep portion of the turbulence spectrum, 
possibly in the damping range beyond the inertial range%
\footnote{Recently, 
a steep turbulence spectrum ($q \simeq 2.7$) in the damping range 
has been adopted by \citet{LiG13} to explain 
the spectral hardening of HXR spectra above 300 keV
in solar flares involving a termination shock.}
\citep{Petrosian10, Petrosian12}.
However, a steep turbulence spectrum will require more energy 
in the turbulence unless its spectral range of the steep part is narrow. 
Obliquely propagating waves will have 
different energy dependences for these timescales,
but limited information on perpendicularly propagating waves 
seems to give similar energy dependences for 
the momentum and pitch angle diffusion timescales \citep{Pryadko99}.

We should emphasize that the two intense events on 2003 November 3
and 2005 September 8 are not representative of typical flares.
As mentioned above, the escape time and energy diffusion time
are primarily determined by the ratio $X_{\rm LT}/X_{\rm FP}$.
The discrepancy between the above observational results
and the SA model predictions is related to 
the relatively bright LT source with a flat spectrum.
On the contrary, 
imaging spectroscopic observations have indicated that
for most flares, the LT source is much weaker and has a steeper spectrum
\citep[e.g.,][]{Petrosian02, LiuW06, Shao09}.
Furthermore, spectral studies of the over-the-limb solar flares
with their FP sources occulted 
from {\it Yohkoh} \citep{Tomczak01, Tomczak09} and
from {\it RHESSI} \citep{Krucker08c, Saint-Hilaire08}
found that their spectral index on average 
is larger than that from the disk flares by 1.5 and 2, respectively.
Thus, for those more common flare events,
we can expect flatter energy dependences for the
escape time, energy diffusion time, and scattering time
as determined from observations,
which would be closer to the SA model predictions.

\subsection{Shock Acceleration}
\label{sec:Shock}

Addition of acceleration by a shock does not seem to help in this regard. 
If a standing shock exists at the acceleration region with 
a high speed $u_{\rm sh}$, %high Alfv\'en Mach number, 
it may dominate the acceleration rate,
\begin{align}
A_{\rm sh}\sim E(u_{\rm sh}/v)^2\langle D_{\mu\mu}\rangle,
\end{align}
with an acceleration time $\tau_{\rm sh}\propto v^2\tau_{\rm scat}$ 
\citep{Petrosian12}. 
If the escape of particles is again diffusive in nature 
($T_{\rm esc} \propto 1/v^2\tau_{\rm scat})$, then 
we expect a simple inverse relation between 
the escape time and shock acceleration time,
$T_{\rm esc} \propto 1/\tau_{\rm sh}$.

On the other hand, if shock acceleration is dominant, 
then from integration of the Fokker-Planck equation (with $D_{\rm EE}=0$), 
we obtain
\begin{align}
A_{\rm sh}(E)= \dot{E}_{\rm L}+ \frac{1}{N}
\int_E^\infty \left(\frac{N}{T_{\rm esc}} - \dot{Q}\right)dE.
\label{eq:DEE3}
\end{align}
Therefore the acceleration rate by a shock,
which now depends only on the pitch angle diffusion rate,
is similar to the energy diffusion coefficient derived above
(Equation \ref{eq:DEE}).
We can then express the shock acceleration timescale 
in terms of observables as
$\tau_{\rm sh}=E/A_{\rm sh}(E)= \tau_{\rm L}(1+X_{\rm FP}/X_{\rm LT})^{-1}$.
For the case $X_{\rm LT} < X_{\rm FP}$, 
our observations indicate that roughly
$\tau_{\rm sh} \propto T_{\rm esc}$ (Equation \ref{eq:timescales}), 
which is in direct disagreement with the above inverse relation
expected from the shock acceleration model.

Thus for both SA and shock acceleration, 
we encounter contradiction with the above Assumption I
that the long escape time is due to 
the random walk approximation expected in 
the strong diffusion limit ($\tau_{\rm scat} \ll \tau_{\rm cross}$). 
Alternatively,
a long escape time may arise in a magnetic mirror geometry 
as we discuss next \citep[e.g.,][]{ChenQ12}.

\subsection{Weak Diffusion and Magnetic Mirroring}

Assumption I involves the use of Equation (\ref{eq:Tesc}),
which in the strong diffusion limit gives the random walk relation,
but in the weak diffusion limit ($\tau_{\rm scat} \gg \tau_{\rm cross}$)
gives $T_{\rm esc}\rightarrow \tau_{\rm cross}$. 
However, in addition to plasma waves or turbulence, 
magnetic reconnection may restructure the 
the large scale magnetic fields into a configuration that
can also trap and accelerate particles in the LT region
\citep[e.g.,][]{Somov97, Karlicky04, Minoshima11, Grady12}.
The newly reconnected, cusp-shaped magnetic field lines 
relax and shrink to the underlying closed loops
and may form a magnetic mirror geometry in the corona. 
A cusp-shaped geometry is often seen from soft X-ray and EUV images and 
coronal HXR sources in many events have been found to be located near 
such a structure \citep[e.g.,][]{Sun12b, LiuW13}. 

If the magnetic field lines converge from the center of the LT acceleration 
region to where the particles escape into the loop legs, 
then the escape time will be affected.
In the strong diffusion limit we would still expect a random walk process.
While in the weak diffusion limit,
instead of $T_{\rm esc}\rightarrow \tau_{\rm cross}$, 
we expect an escape time $T_{\rm esc}\propto \tau_{\rm scat}$, 
which is the time needed to scatter particles into the loss cone
\citep{Kennel69, Melrose76}. The proportionality constant 
depends on several factors but primarily on the pitch angle distribution 
and the mirroring ratio $\eta_m \sim B_{\rm L}/B_0$,
the ratio of the magnetic field intensity from 
the boundary to the center of the acceleration region. 
As well known from the conservation of the magnetic moment
(the first adiabatic invariant),
only particles with a pitch angle smaller than the mirroring angle, 
or a pitch angle cosine $|\mu| > \mu_{\rm cr}=\sqrt{1-1/\eta_m}$, 
can escape from this magnetic trap and penetrate to the loop FPs.
In absence of scattering,
the rest of the particles will be trapped in the mirror.
But when there is scattering, 
the particles with high pitch angles will be scattered into the loss cone, 
perhaps after several bounces back and forth between the mirroring points. 
For example, for an isotropic pitch angle distribution, 
we can obtain an average escape time as
\begin{equation}
\frac{1}{T_{\rm esc}}\sim \frac{1-\mu_{\rm cr}}{\tau_{\rm cross}}+
\frac{\mu_{\rm cr}}{\tau_{\rm scat}},
\end{equation}
which for strong convergence $(\mu_{\rm cr}\rightarrow 1$) 
would give $T_{\rm esc} \sim \tau_{\rm scat}$.
\citet{Malyshkin01} showed that
for electrons injected into the trap with $\mu=0$,
one has a proportionality constant of $\ln \eta_m$.
For strong convergence, they also identified an intermediate range 
$1/2\eta_m\ll \tau_{\rm scat}/\tau_{\rm cross}\ll 2\eta_m$, 
where $T_{\rm esc}\sim 2\eta_m \tau_{\rm cross}$. 

In summary, in a converging field geometry, we expect 
the escape time first decreases with an increasing scattering time, 
but instead of becoming equal to the crossing time, 
it then reaches a minimum and 
begins to increase linearly with the scattering time 
when the latter exceeds the crossing time. 
Thus, a long escape time can arise not only from a short scattering time
in the strong diffusion scenario, but also from 
a long scattering time in a converging magnetic field configuration.
Some earliest mechanisms that were proposed to produce 
a distinct coronal HXR source \citep[e.g.,][]{Leach84, Fletcher98}
basically adopted the second scenario with Coulomb collisions 
as the scattering agent for the suprathermal electrons.
In addition, an observed escape time that increases with energy requires 
a scattering time that also increases with energy.
For scattering due to Coulomb collisions 
in a fully ionized hydrogen plasma in the non-relativistic limit,
the electron pitch angle scattering time follows
$\tau_{\rm scat}^{\rm Coul}\sim \tau_{\rm L}^{\rm Coul} \propto E^{3/2}$
\citep{Trubnikov65, Melrose76, Bai82, Aschwanden02}.

Therefore, if Coulomb scattering is the agent that scatters electrons
into the loss cone and cause their escape,
the escape time will be comparable to the energy loss time and 
scales with energy as $T_{\rm esc} \propto E^{3/2}$. 
This is in disagreement with the above observational results 
from both events. 
Even if the scattering time is comparable to the crossing time, 
then we would be in the intermediate regime and the escape time 
would vary like the crossing time as $E^{-1/2}$, 
which also disagrees with observations.

On the other hand, 
if scattering is dominated by wave-particle interactions, 
then because the above observations gives roughly
similar energy dependences for the energy diffusion time and escape time, 
this would mean similar dependences for
the scattering time and energy diffusion time.
This would be in better agreement with theoretical expectations
in the SA model by turbulence. 
But for the shock acceleration model, 
our observations imply a relation $T_{\rm esc}\propto \tau_{\rm sh}$,
and in a converging field configuration, 
we expect $T_{\rm esc}\sim \tau_{\rm scat}$.
While the shock acceleration model
predicts $\tau_{\rm sh}\propto v^2 \tau_{\rm scat}$
(Section \ref{sec:Shock}).

We therefore conclude that this very first non-parametric determination 
of the SA model characteristics directly from the observed data 
could be reconciled with stochastic acceleration by turbulence,
if the LT acceleration region is surrounded by 
a cusp-shaped magnetic geometry with a relatively large mirroring ratio. 
On the other hand, our results do not 
seem to be consistent with what is expected in acceleration 
in a standing shock at the LT region of the flare, 
regardless of whether there is strong field convergence.
The exact treatment of the acceleration and transport of electrons in 
a more realistic geometry cannot be treated by the leaky box model and 
requires inclusion of the kinematic effects of a magnetic mirror 
in a dynamic flare environment that is more complicated 
and will be treated in future works.

\acknowledgments

This work was supported by NASA grants NNX10AC06G and NNX13AF79G. 
Q.C. thanks the discussions and help 
from Anna Massone with the electron flux spectral images
and Kim Tolbert with OSPEX.
We thank the referee for constructive comments.
{\it RHESSI} is a NASA small explorer mission. 

{\it Facilities:} \facility{{\it RHESSI}}.

\end{document}